\documentclass[a4paper,11pt]{article}
\usepackage{amsmath}
\usepackage{mathrsfs}

\providecommand{\dd}[2][]{\frac{d #1}{d #2}}
\title{Parisi-Wu Quantization, CP violation and Beyond the Standard Model}

\author{A. K. Kapoor\footnote{Former Adjunct Professor, akkapoor@cmi.ac.in}\\ Chennai Mathematical Institute\\
H1 SIPCOT, IT Park, Siruseri\\
Kelambakkam, TN 603103, INDIA}

\begin{document}

\maketitle
\vfill
\begin{center}
 (to appear in Modern Physics Letters A)
\end{center}

\vfill
\noindent
Keywords:
Anomalous axial vector gauge model; Parisi-Wu quantization; Operator formalism; Ultraviolet divergences; Chiral symmetry; Stochastic supersymmetry; CP Violation; Beyond standard model.\\[5mm]

\newpage
\begin{abstract}
The Parisi-Wu scheme of quantization opens up the possibility of using anomalous fermionic gauge theories. An analysis of ultra-violet divergences reveals that the structure of counter terms  is different from what is expected in conventional quantization schemes. 
In this letter it is argued that there exists a possible mechanism of CP violation that requires at least three generations of quarks, a result well known from a phenomenological analysis of 
mass mixing of quarks.  A few observations on possible ways of going beyond the standard model are included.
\end{abstract}

The Parisi-Wu scheme of stochastic quantization was proposed as an alternative scheme of quantization having an advantage of not requiring conventional gauge fixing. For a detailed account of this scheme see 
\cite{SQM1, SQM2, SQM3, SQM4}.
In this letter we make a few observations on the Parisi-Wu stochastic quantization scheme of a class of anomalous fermionic  gauge theories. 
Simplest model in this class is a single fermion coupled to an axial vector gauge field. 

There exists several different ways of formulating the stochastic quantization method (SQM). For example we have the Langevin formalism, the Fokker-Planck
formalism and the operator formalism pioneered by Namiki and Yamanaka \cite{MNYY}. 

Many formal proofs have been given to demonstrate  that the stochastic quantization scheme is equivalent to the conventional quantization schemes (CQS). An important symmetry, called stochastic super symmetry in SQM, ensures this equivalence of SQM and conventional quantization schemes \cite {Pramana}.

\paragraph*{A brief overview of SQM}
 In the Parisi-Wu SQM, the fields are assumed to be  stochastic processes in a fictitious time, \(t\), also called the stochastic time. In the equilibrium limit, the SQM formally becomes equivalent to CQS. 
 
 In the operator formalism, most useful for our discussion, the SQM formalism becomes quantum field theory in five dimensions. It is the most convenient way to carry out the analysis of ultra violet divergences.  Apart from the original fields \(\phi_k\), one needs to introduce a stochastic momentum field \(\pi_k\) corresponding to every field in the model.
 The stochastic action \(\Lambda \) of the five dimensional field theory of the operator formalism is constructed in terms of Euclidean action \(S_E[\phi_k]\) as follows.
 \begin{eqnarray}
  S_E &=& \int d^4 x \mathscr L\\
  \Lambda = \int d^4x dt\Big[\pi_k \dd[\phi_k]{t} - \mathcal H\Big], \qquad\qquad
  \mathcal H &=& \gamma^{-1} \pi_k\pi_k - \phi_k\frac{\delta S_E}{\delta \phi_k}
 \end{eqnarray}
 The object \(\mathcal H\) is known as the stochastic Hamiltonian and the expression for \(\mathcal H\) gets  its form from the Fokker-Planck  equation. 
 
\section*{SQM of U(1) anomalous gauge model}
 This model has been subject of investigation in a set of earlier papers \cite{akk15,MPLA19,akk20, akk21, akk21A}. For stochastic quantization of fermionic theories see \cite{Sand}. Here we will give, for reference, a few basic details  of  the fermion model with \(U(1)\) gauged chiral symmetry. More details can be found in \cite{akk21}. For SQM of fermionic theories  see, for example, \cite{Sand}.

\(U(1)\) axial vector gauge theory will be taken to be  
\begin{eqnarray}
   S_E &=& \int d^4 x \mathscr{L} \\
   \mathscr{L} &=&  \bar{\psi}(-i\gamma_\mu D_\mu + m)
\psi - \frac{1}{4} F_{\mu\nu}F_{\mu\nu} + \frac{M^2}{2}A_\mu A_\mu  -
\frac{1}{2\alpha}(\partial\cdot A)^2
\end{eqnarray}
where
\begin{eqnarray}
 D_\mu = \partial_\mu -ig\gamma^5 A_\mu,\qquad F_{\mu\nu} = \partial_\mu A_\nu-
\partial_\nu A_\mu,
\end{eqnarray}
\subsubsection*{\it Operator formalism}
Letting \(\pi_\mu,
\bar{\omega},\omega\) to denote the stochastic momenta corresponding to
the gauge field \(A_\mu\) and the fermionic fields \(\psi, \bar{\psi}\),
the stochastic action \(\Lambda\) of the five dimensional field theory takes
the following form:
\begin{equation}
 \Lambda = \int dx d\tau\left( \pi_\mu \frac{\partial A_\mu}{\partial
t}  + \frac{\partial\bar{\psi}}{\partial t}\omega  + \bar{\omega}
\frac{\partial\psi}{\partial t} - {\mathcal H}
 \right) \label{EQ07},
\end{equation}
where
\begin{eqnarray}
{\mathcal H}&=& \left[ \gamma^{-1} \pi_\mu\pi_\mu + 2\bar{\omega}K \omega
-\bar{\omega}\tilde{K}\frac{\delta S_E}{\delta \bar{\psi} } + \frac{\delta
S_E}{\delta \psi}\tilde{K} \omega -
\gamma^{-1}
\pi_\mu\frac{\delta S_E}{\delta A_\mu}
\right] \label{EQ08}.
\end{eqnarray}

Here \(K(x,t)\) is a suitable kernel that  needs to be used for the fermions.
For our present discussion,  an explicit expression for the kernel for fermions
is not required. See \cite{MPLA19} for more details.

A term
\(\frac{1}{2\alpha}(\partial_\mu A_\mu)^2\) has been included so that the
degree of divergence remains bounded.

\subsubsection*{\it Ultra violet divergences and counter terms}
A study of ultra violet divergences of the stochastic field theory defined by \eqref{EQ07} reveals that the finiteness of the theory requires addition of a new counter term 
\begin{equation}\label{EQ081}
\mathscr{L}_{\pi-A-A} = f\epsilon_{\mu\nu\alpha\beta} \pi_\mu (\partial_\nu A_\alpha) A_\beta
\end{equation}
Such a term is not present in the starting stochastic Hamiltonian \(\mathcal{H}\) of Eq \eqref{EQ08}. 

In previous paper \cite{MPLA19}, we have  obtained a Ward identity which
ensures that unphysical degree of freedom, the longitudinal  component of the
axial vector  field, decouples if  a suitable choice of the coupling constant \(f\) is made.

The straightforward investigation of ultra violet divergences, by doing a power counting, mentioned above showed presence of a, new additional term as in \eqref{EQ081}. This term corresponds to proper vertex diagram with three external lines corresponding to the \(\pi_\mu A_\nu, A_\sigma\) fields.  The coefficient of this term is a free parameter and can be chosen to cancel the anomalous term in the Ward identities involving the gauge field. The appearance of this counter term has far reaching consequences that need to be explored.
\paragraph*{\it Decoupling of unphysical degrees of freedom}
In the CQS, the fermion axial vector current is not conserved due to anomaly. Since the  
the gauge field couples to the non-conserved, fermion axial vector current, the longitudinal degrees of freedom,\(\partial_\mu A_\mu\), of the gauge field does not decouple and the unitarity does not hold.

In SQM, the fermion axial vector current remains anomalous. But the current coupled to the axial gauge field receives an extra contribution from the new counter term, that makes it a conserved current. Thus the unphysical degrees of freedom of the axial gauge field decouple making the 
theory unitary.

\paragraph*{Chiral symmetry} The stochastic action breaks the chiral symmetry, this fact raises the possibility of generation of no zero fermion mass by radiative corrections.

\paragraph{Stochastic supersymmetry}
It has been mentioned in one of the opening paragraphs that  that a stochastic supersymmetry ensures equivalence of SQm with CQS of a field theory model.
The stochastic supersymmetry is broken in presence of the \(\pi-A-A\) counter term. Thus there does not exist a  quantum field theory in CQS which is the equilibrium limit of the field theory of SQM operator formalism.

\paragraph*{CPT theorem}
Since the SQM of anomalous fermion theory is not equivalent to a local Lagrangian field theory in four dimensions, the SQM version is expected to break CPT symmetry. Thus one may have  violations of corresponding results that follow from CPT theorem.

\paragraph*{CP Violation} The investigations in previous papers has been limited to the abelian axial vector gauge field coupled to a fermion.
One may try guessing what happens in we look at the non-abelian anomalous gauge theory with fermions. It is expected that unphysical gauge degrees of freedom will decouple as in the abelian case. In this case  the counter term for the three point vertex function  will have the form \(\epsilon_{\mu\nu\alpha\beta} D_\mu \pi_\nu  F_{\alpha\beta}\) where \(D_\mu\) is  a covariant derivative. Here it is important to note that the \(U(1)\) chiral gauge model is too simple and is unlikely to be of any use for going beyond the standard model. One need not restrict the gauge field fermion coupling to be of the axial vector type. Depending on the coupling to the fermions, this term can give rise to CP violation.

\paragraph*{Why three generations of quarks and leptons?} We ask what is the simplest choice of the gauge group that may give rise to CP violation via the mechanism mentioned above?
An axial vector gauge theory based on  \(SU(2)\) gauge group is known to be anomaly free and does not require any further discussion in the presence context. The next unitary group \(SU(3)\) with axial vector gauge fields does not automatically lead to anomaly free theory. With fermions belonging  three dimensional representation, the theory will, in general, be anomalous and CP violation.

Thus we have an important  prediction that a model of CP violation based on SQM of anomalous theories requires at least three generations of quarks and leptons. This conclusion is  already known from the analysis of models in which the source of  CP violation is the  mixing of quark fields coming from the mass matrix.  

\paragraph*{A possible path to go beyond the standard model}
Experimentally, we have three generation of leptons and three generations of quarks. 
The discussion related to CP violation in the previous paragraph suggests that a  \(SU(3)\)  anomalous gauge theory  be investigated, the group  transformations may be chosen to be that acting on the generation index of the quarks and leptons. It will be noticed  that gauge fields, in this model, will all be neutral and may as well be candidates for dark matter. 

To summarize, it has been argued that the SQM of anomalous theories opens up a whole new class of models suited to go beyond the standard electro-weak model. This class of models is crying for a  need to sit up and investigate.

\end{document}